\documentclass[journal=jacsat,manuscript=article]{achemso}

\usepackage[version=3]{mhchem}

\usepackage{placeins}
\usepackage{booktabs}
\usepackage{multirow}
\usepackage{arydshln}
\usepackage{threeparttable}
\usepackage{hyperref}

\usepackage[T1]{fontenc}
\usepackage{amsmath}
\newcommand{\DMS}{\mathord{\textit{\DJ}}}

\author{Jens Wagner}
\author{Thomas Specht}
\author{Hans Hasse}
\author{Fabian Jirasek}
\email{fabian.jirasek@rptu.de}
\phone{+49 (0)631 - 205 4685}
\affiliation[Unknown University]
{Laboratory of Engineering Thermodynamics (LTD), RPTU Kaiserslautern, Germany}

\title[An \textsf{achemso} demo]
  {Hybrid Machine Learning for Physically Consistent Prediction of Mutual Diffusion Coefficients in Binary Liquid Mixtures}

\begin{document}

\begin{abstract}

Experimental data on mutual diffusion coefficients in liquid mixtures are scarce, and existing prediction methods are unreliable for non-ideal mixtures.
Building on HADES, our hybrid model for self-diffusion coefficients, we introduce the Enhanced Vignes Equation~(EVE), which augments the Vignes equation with a deep-set model. 
Its inputs are molecular descriptors derived from SMILES and activity-coefficient information from HANNA, a machine-learning model for activity coefficients.
EVE predicts Maxwell--Stefan diffusion coefficients; the corresponding Fickian coefficients follow from the thermodynamic factor, likewise obtained from HANNA.
Trained and evaluated on 2106 experimental data points for 123 binary mixtures, EVE outperforms all available semi-empirical models, halving the mean absolute relative error for mixtures without experimental infinite-dilution diffusion data.
Combined with HADES, EVE yields the complete set of self- and mutual diffusion coefficients of a binary mixture from SMILES and pure-component viscosities alone.
EVE and HADES are available at \href{https://ml-prop.mv.rptu.de/}{\mbox{https://ml-prop.mv.rptu.de/}}, and the code is fully disclosed.

\end{abstract}

\section{Introduction}

Mutual diffusion coefficients in liquid mixtures are needed to describe mass transfer in many natural and industrial systems.
In particular, they are central to designing and optimizing separation equipment and reactors~\cite{Wesselingh2000, Bellaire2022b}.
However, their experimental determination is laborious, so that available data cover only a small fraction of technically relevant mixtures.
Mutual diffusion can be described either in the Fickian or in the Maxwell–Stefan picture.
The two descriptions are thermodynamically equivalent and are linked by the thermodynamic factor.
For predictive modeling, the Maxwell--Stefan formulation is usually preferred, as the Maxwell--Stefan diffusion coefficient~$\DMS_{ij}$ depends less strongly on composition than its Fickian counterpart~$D_{ij}$~\cite{Rehfeldt2010}.
In the limit of infinite dilution of component~$i$ in component~$j$, the Fickian and Maxwell–Stefan mutual diffusion coefficients and the self-diffusion coefficient of $i$ all coincide; we therefore refer to this common value simply as the diffusion coefficient at infinite dilution, $D^\infty_{ij}$.

While numerous methods are available for predicting diffusion coefficients at infinite dilution~\cite{Taylor1993, Wilke1955, Reddy1967, Tyn1975, Evans2013, Evans2018, Khajeh2011, Abbasi2014, Mariani2020, Aniceto2021, Aniceto2024, Dias2024, Gromann2022, Romero2025, Romero2026, Wagner2026b}, far fewer address the composition dependence of mutual diffusion coefficients in mixtures~\cite{Darken1948, Vignes1966, Krishna2005, Carman1956, Leffler1970, Li2001, Bosse2006, Rehfeldt2007, DAgostino2011, Zhou2013, Zhu2015,Beigzadeh2012, Abbasi2014}.
The established approach for binary mixtures is the Vignes equation~\cite{Vignes1966}, which interpolates between the two infinite-dilution coefficients $D^\infty_{ij}$ and $D^\infty_{ji}$ to obtain $\DMS_{ij}$ at finite concentrations.
The generalized Darken model~\cite{Darken1948, Krishna2005} offers an alternative that additionally uses self-diffusion coefficients of the pure components as input.
However, because both methods are essentially interpolation schemes, they cannot capture the composition dependence observed in many real mixtures.
Numerous empirical extensions have therefore been proposed~\cite{Carman1956, Leffler1970, Li2001, Bosse2006, DAgostino2011, Zhou2013, Zhu2015}, introducing mixture-specific parameters or correction factors fitted to experimental data.
Their accuracy nonetheless remains limited even for weakly non-ideal mixtures~\cite{Rehfeldt2010, Parez2013, GuevaraCarrion2016}, reflecting a lack of robustness outside the cases for which they were developed.
Their applicability is further restricted by data scarcity: self-diffusion coefficients at finite concentrations, which several of these models require as input, are even rarer than mutual diffusion coefficients themselves~\cite{Gromann2022}.

Data-driven machine learning~(ML) models provide an alternative.
However, only two such models have been developed so far to predict mutual diffusion coefficients at finite concentrations~\cite{Beigzadeh2012, Abbasi2014}, one of which is restricted to hydrocarbon mixtures~\cite{Abbasi2014}.
Although both achieved higher accuracy than semi-empirical methods, their fully data-driven nature does not enforce physical constraints, such as a smooth dependence of the diffusion coefficients on composition or invariance of the predictions with respect to permutation of the mixture components.
Nothing in these models prevents such violations, which can therefore occur at any state point and go unnoticed by the user.

What is therefore lacking is a model that combines the accuracy of data-driven methods with the consistency of physical ones, without relying on experimental input that is largely unavailable.
Hybrid models, which combine physical models with machine-learning architectures, offer a promising route to this goal, as they retain the consistency and interpretability of the physical basis while exploiting ML flexibility~\cite {Jirasek2020b, Jirasek2021, Jirasek2023b, Hoffmann2025c, Hasse2026, Romero2026b}.
We have recently demonstrated this for self-diffusion by introducing the HADES model~\cite{Wagner2026c}, which combines the Stokes--Einstein equation with a deep-set model~(DSM)~\cite{Zaheer2017}, a neural network whose output does not depend on the order of its inputs, to predict self-diffusion coefficients in mixtures with any number of components across the full composition range, requiring only the SMILES~\cite{Weininger1988} strings of the components and their pure-component viscosities as input.
HADES also predicts the diffusion coefficients at infinite dilution~$D^\infty_{ij}$, and thus provides precisely the quantities on which Vignes-type models depend.

In this work, we extend this hybrid approach to mutual diffusion by augmenting the Vignes equation with a DSM to yield the Enhanced Vignes Equation~(EVE).
Besides the infinite-dilution coefficients entering the Vignes part, EVE requires only the SMILES strings of the components, from which molecular descriptors are generated, together with activity-coefficient information predicted by HANNA~\cite{Specht2024, Hoffmann2025b}, our ML model for activity coefficients.
We can use infinite-dilution diffusion coefficients from experiments where available or predict them with HADES so that EVE can be applied either way; we benchmark both scenarios against the Vignes equation and other available semi-empirical models.
Finally, we show that HADES and EVE together accurately predict the complete set of self- and mutual diffusion coefficients of previously unseen binary mixtures, without requiring any experimental diffusion data as input.

\section{Methods}

\subsection{Theoretical Background}
\label{sec_tb}

We consider binary mixtures of components $i$ and $j$ throughout.
Mutual diffusion in such a mixture is described by a single diffusion coefficient, which can be expressed either in the Fickian picture, where the flux relates to the concentration gradient, or in the Maxwell–Stefan picture, where it relates to the chemical potential gradient.
The corresponding diffusion coefficients $D_{ij}$ and $\DMS_{ij}$ are symmetric, $D_{ij} = D_{ji}$ and $\DMS_{ij} = \DMS_{ji}$, and are related by
\begin{equation}
D_{ij} = \DMS_{ij} \Gamma_{ij},
\label{eq_FMS}
\end{equation}
where $\Gamma_{ij}$ is the thermodynamic factor.
With $\gamma_i = \gamma_i \left( T, x_i \right)$ denoting the activity coefficient of component~$i$ at temperature~$T$ and mole fraction~$x_i$, it is given by
\begin{equation}
\Gamma_{ij} = 1 + x_i \frac{\partial \ln \gamma_i }{\partial x_i}.
\label{eq_TCF}
\end{equation}
In the limit $x_i \to 0$, the thermodynamic factor approaches unity, and $D_{ij}$, $\DMS_{ij}$, and the self-diffusion coefficient of $i$ all coincide.
This common value is the diffusion coefficient at infinite dilution, $D^\infty_{ij}$.
In contrast to $D_{ij}$ and $\DMS_{ij}$, it is not symmetric: $D^\infty_{ij}$ and $D^\infty_{ji}$ are the coefficients of $i$ infinitely diluted in $j$ and of $j$ infinitely diluted in $i$, respectively, and generally differ.
The Vignes equation~\cite{Vignes1966} predicts the composition dependence of $\DMS_{ij}$ from these two coefficients:
\begin{equation}
\DMS^\mathrm{Vig}_{ij} = \left( D_{ij}^{\infty} \right)^{x_j} \left( D_{ji}^{\infty} \right)^{x_i}.
\label{eq_Vig}
\end{equation}

\subsection{The EVE Model}
\label{sec_eve}

\begin{figure}
\centering
    \includegraphics[width=14.5cm]{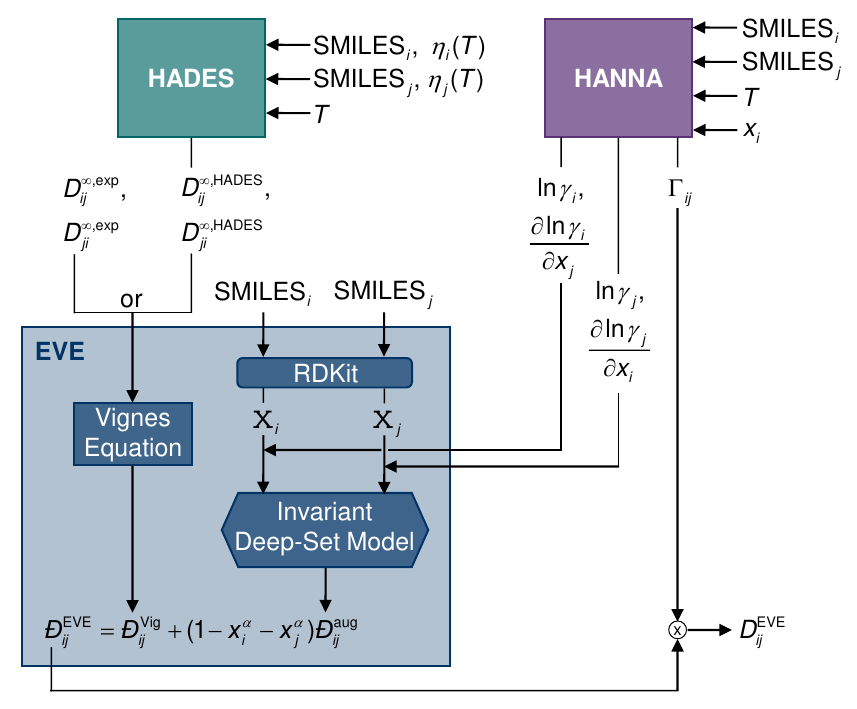}
    \caption{Schematic overview of the Enhanced Vignes Equation~(EVE), which combines the Vignes equation with an invariant deep-set model~(DSM)~\cite{Zaheer2017}, and of the prediction framework, incorporating HADES~\cite{Wagner2026c} and HANNA~\cite{Specht2024}, for the physically consistent prediction of concentration-dependent Maxwell--Stefan and Fickian diffusion coefficients,~$\DMS_{ij}^\mathrm{EVE}$ and $D_{ij}^\mathrm{EVE}$, respectively. Technical details on the EVE architecture and the prediction framework are provided in Sections~\nameref{sec_eve} and \nameref{sec_pfw}, respectively.}
    \label{fig_EVE}
\end{figure}

Figure~\ref{fig_EVE} shows the EVE model architecture.
Its physical core is the Vignes equation, Eq.~\eqref{eq_Vig}, which yields a preliminary prediction $\DMS^\mathrm{Vig}_{ij}$ of the Maxwell--Stefan diffusion coefficient from the two infinite-dilution diffusion coefficients $D^\infty_{ij}$ and $D^\infty_{ji}$ of the considered binary mixture.
This prediction is then augmented by a deep-set model~(DSM)~\cite{Zaheer2017} giving the EVE prediction
\begin{equation}
\DMS_{ij}^\mathrm{EVE} = \DMS_{ij}^\mathrm{Vig} + \left( 1 - x_i^\alpha - x_j^\alpha \right) \DMS_{ij}^\mathrm{aug},
\label{eq_EVE}
\end{equation}
where $\DMS_{ij}^\mathrm{aug}$ is the augmentation produced by the DSM and $\alpha$ is a hyperparameter.
The prefactor $\left( 1 - x_i^\alpha - x_j^\alpha \right)$ vanishes in the limits $x_i \to 0$ and $x_i \to 1$, so that $\DMS_{ij}^\mathrm{EVE}$ reduces to $\DMS_{ij}^\mathrm{Vig}$ there.
EVE therefore reproduces the diffusion coefficients at infinite dilution exactly, regardless of the augmentation.
Several functional forms of a prefactor with this property were tested in preliminary studies; the one used in Eq.~\eqref{eq_EVE} performed best and was therefore adopted here.
Its hyperparameter $\alpha$ determines how rapidly the augmentation is suppressed as the pure components are approached: the larger $\alpha$, the closer to the boundaries the DSM remains free to modify the Vignes prediction.
In this work, we set $\alpha$ to 12, as determined by hyperparameter optimization (see below).
Because the prefactor is symmetric in $x_i$ and $x_j$, Eq.~\eqref{eq_EVE} is invariant under permutation of the components.

\FloatBarrier

The augmentation in Eq.~\eqref{eq_EVE} is additive and unbounded, so $\DMS_{ij}^\mathrm{EVE}$ is not guaranteed to be positive by construction.
However, for none of the data considered in this work did EVE predict a negative value.
Nevertheless, predictions are truncated at the machine epsilon at inference, which serves as a safeguard for the user.

For each component, the DSM inputs are a vector of molecular descriptors $\mathbf{X}$ and activity-coefficient information.
We generate the descriptors automatically from the SMILES~\cite{Weininger1988} strings using RDKit~\cite{RDKit2024}; the descriptor set, given in Table~\ref{tab_x}, matches that in our previous work on HADES~\cite{Wagner2026c}, keeping the two models consistent and avoiding an untested representation.
The activity-coefficient information includes the logarithmic activity coefficient and its derivative with respect to the other component's mole fraction.
Including this information on the mixture's thermodynamic non-ideality, both its magnitude and its local change with composition, is essential for the DSM to capture the composition dependence of $\DMS_{ij}$.
\begin{table}
\centering
\caption{Molecular descriptors included in the vectors~$\mathbf{X}_i$ and $\mathbf{X}_j$ describing the mixture components~$i$ and $j$, respectively, as used in the EVE model.}
\label{tab_x}
    \begin{tabular}{ll}
        \toprule
        Label & Molecular descriptor \\
        \midrule
        $M$ & Molar mass in kg~mol$^{-1}$\\
        $R$ & Boolean variable indicating presence of molecular ring structures\\
        $r_\mathrm{Het}$ & Ratio of number of heteroatoms to non-hydrogen atoms\\
        $r_\mathrm{Hal}$ & Ratio of number of halogen atoms to non-hydrogen atoms\\
        $r_\mathrm{Acc}$ & Ratio of number of hydrogen-bond acceptors to non-hydrogen atoms\\
        $r_\mathrm{Don}$ & Ratio of number of hydrogen-bond donors to non-hydrogen atoms\\
        \bottomrule
    \end{tabular}
\end{table}
\FloatBarrier

The DSM processes the inputs for the two components in parallel and aggregates them symmetrically, so that $\DMS_{ij}^\mathrm{aug}$, and hence $\DMS_{ij}^\mathrm{EVE}$, does not depend on the order in which the components are supplied~\cite{Soelch2019, Wagstaff2022}.
Temperature is not an input to the DSM.
It enters EVE through two channels: the infinite-dilution diffusion coefficients in the Vignes part and the activity-coefficient information supplied to the DSM.
Since the infinite-dilution diffusion coefficients generally increase with temperature, so does $\DMS_{ij}^\mathrm{Vig}$ at any fixed composition.
The activity-coefficient information, in contrast, can change with temperature in either direction, and its influence on $\DMS_{ij}^\mathrm{aug}$ is not constrained.
Monotonicity of $\DMS_{ij}^\mathrm{EVE}$ with respect to temperature is therefore not enforced by construction.

Hyperparameter optimization (see below) yielded a DSM architecture consisting of an input layer, one deep set layer~\cite{Zaheer2017}, and an output layer, each with 32 nodes and a Sigmoid Linear Unit (SiLU) activation function, with sum aggregation following the deep set layer~\cite {Soelch2019}.

\subsection{Prediction Framework}
\label{sec_pfw}

The Vignes part of EVE requires infinite-dilution diffusion coefficients, and its DSM requires activity-coefficient information; both must be supplied from outside the model.
Figure~\ref{fig_EVE} shows the framework used for this purpose in the present work.

For the infinite-dilution diffusion coefficients, experimental values $D^{\infty,\mathrm{exp}}_{ij}$ and $D^{\infty,\mathrm{exp}}_{ji}$ are used where available; where they are not, they are predicted with HADES~\cite{Wagner2026c}, which requires the SMILES strings of the two components, their pure-component viscosities~$\eta(T)$, and the temperature~$T$.

The activity coefficients and their composition derivatives are obtained from HANNA~\cite{Specht2024}, which predicts them from the components' SMILES strings, temperature, and composition.
HANNA also provides the thermodynamic factor $\Gamma_{ij}$, from which the Fickian diffusion coefficient follows as $D^\mathrm{EVE}_{ij} = \DMS^\mathrm{EVE}_{ij} \Gamma_{ij}$, cf. Eq.~\eqref{eq_FMS}.
Details on the use of HANNA, including the computation of the composition derivatives, are given in the Supporting Information.

HADES and HANNA were chosen because they are the state of the art for the physically consistent prediction of infinite-dilution diffusion coefficients and activity coefficients, respectively, and because both require only SMILES strings and, for HADES, pure-component viscosities.
The complete framework therefore predicts the Maxwell--Stefan and Fickian mutual diffusion coefficients of a binary mixture at any temperature and composition from the SMILES strings of its components and their pure-component viscosities.
Neither model is intrinsically required by EVE: any other source of consistent infinite-dilution diffusion coefficients and activity coefficients, including experimental data, can be used instead.

\subsection{Experimental Database}

As in our work on HADES~\cite{Wagner2026c}, we consider liquid mixtures of non-ionic organic components (including water) with molar masses up to 1000 g mol$^{-1}$ and containing no atoms heavier than chlorine.
This restriction reflects the availability of experimental data rather than a limitation of the modeling approach.

All experimental Fickian diffusion coefficients $D^\mathrm{exp}_{ij}$ for binary mixtures were retrieved from the Dortmund Data Bank (DDB)~\cite{DDB2025}.
Obtaining the corresponding Maxwell--Stefan diffusion coefficients from these data requires a thermodynamic model; the values $\DMS^\mathrm{exp}_{ij}$ used here were determined from the measured Fickian coefficients using the thermodynamic factors predicted by HANNA, cf. Eq.~\eqref{eq_FMS}.
Data points for which this conversion did not yield physically meaningful results were discarded; the criteria and the number of points affected are given in the Supporting Information, together with a comparison of the resulting $\DMS^\mathrm{exp}_{ij}$ with literature values obtained using conventional activity-coefficient models, which shows good agreement.

The resulting database comprises $N = 2106$ pairs of $D^\mathrm{exp}_{ij}$ and $\DMS^\mathrm{exp}_{ij}$ for 123~binary systems composed of 66~components, covering temperatures from 273~K to 348~K.
A list of all components with their SMILES strings is given in the Supporting Information.

The experimental diffusion coefficients at infinite dilution, $D^{\infty,\mathrm{exp}}_{ij}$ and $D^{\infty,\mathrm{exp}}_{ji}$, used in EVE and in the reference models (see below), and the self-diffusion coefficients $D^\mathrm{exp}_{i}$, used for the comparison with HADES predictions, were taken from the database compiled in our previous work~\cite{Wagner2026c}.
The pure-component viscosities required by HADES were taken from the DDB~\cite{DDB2025}.

\subsection{Training and Evaluation}
\label{sec_tae}

We trained and evaluated the EVE model using a $K$-fold cross-validation~(CV)~\cite{Stone1974} based on system-wise data splits, where $K = 123$ equals the number of distinct binary systems in our database and, in each fold, all data associated with one system were held out as test data, irrespective of composition and temperature.
We then randomly split the remaining data point-wise into training (80\%) and validation (20\%) sets, and repeated this procedure for all folds, yielding a prediction for every data point in the database from a model not trained on any mutual diffusion coefficient of the corresponding mixture.

The Vignes contribution~$\DMS_{ij}^{\mathrm{Vig}}$ in EVE was obtained using experimental infinite-dilution diffusion coefficients where available, and HADES predictions where experimental data were missing.
We used HADES and HANNA, the latter supplying the activity-coefficient information, as published and therefore did not retrain them for each fold; the infinite-dilution coefficients entering the Vignes contribution may thus stem from a model that has seen data for the system concerned.
The CV results therefore assess EVE's predictive capability, given the input quality the framework provides in practical application.

During training, the weights of the DSM were optimized by minimizing the mean squared relative error~(MSRE), which served as the loss function and is defined as the average, over all considered data points, of the squared relative error~(SRE) between the predicted and experimental values:
\begin{equation} \mathrm{SRE} = \left( \frac{\DMS_{ij}^{\mathrm{pred}} - \DMS_{ij}^{\mathrm{exp}}}{\DMS_{ij}^{\mathrm{exp}}} \right)^2 .
\label{eq_SRE}
\end{equation}

The weights were updated using the AdamW~optimizer~\cite{Loshchilov2017}, with the learning rate reduced by a factor of 0.1 after 25~epochs without improvement in the validation loss.
Training was stopped when the validation loss had not improved for 50~consecutive epochs, and the model with the lowest validation loss was retained for evaluation on the test set.

Hyperparameter optimization was performed using a grid-search approach based on the validation loss.
The resulting hyperparameters, along with a sensitivity analysis and the associated validation-loss results, are provided in the Supporting Information.

In addition to SRE and MSRE, model performance was further evaluated using the absolute relative error (ARE), defined as
\begin{equation} \mathrm{ARE} = \left| \frac{\DMS_{ij}^{\mathrm{pred}} - \DMS_{ij}^{\mathrm{exp}}}{\DMS_{ij}^{\mathrm{exp}}} \right|.
\label{eq_ARE}
\end{equation}

Beyond the above-described mixture-wise CV, a stricter test is applied in the final part of the \nameref{sec_rad} section, where the complete set of diffusion coefficients is predicted for binary systems for which experimental data on all diffusion coefficients are available.
There, EVE and HADES were each retrained on their complete databases except for all data on the system concerned, using the training procedures and hyperparameters of this work and of the original work~\cite{Wagner2026c}, respectively.
These predictions therefore involve no experimental diffusion data for the considered system at any stage.

While the evaluations above assess predictive capability, we also provide a 'final' version of EVE.
This final version consists of an ensemble~\cite{Dietterich2000} of ten models, trained by generating ten independent random point-wise splits of the full database into training (95\%) and validation (5\%) sets.
For each split, we initialized a separate model with a distinct random seed and trained it independently.
The ensemble prediction is computed as the arithmetic mean of the individual model predictions.
The final EVE ensemble is available on Zenodo (\href{https://doi.org/10.5281/zenodo.22848728}{\mbox{https://doi.org/10.5281/zenodo.22848728}}) and via the interactive website MLPROP~\cite{Hoffmann2025} (\href{https://ml-prop.mv.rptu.de/}{\mbox{https://ml-prop.mv.rptu.de/}}).

We trained the model on a CPU node in the HPC Elwetrisch (dual Intel Xeon Silver 4410T, 64 GB RAM), using a single CPU core.
We implemented all models, training, and evaluation routines in Python~3.12.7 using PyTorch~2.2.1~\cite{Paszke2019}.
Typical training times per fold ranged from 3 to 5~min.

\subsection{Use of Artificial Intelligence Tools}
During the preparation of this work, the authors used Claude~(Anthropic) to rephrase and condense parts of the manuscript text and to assist in writing and debugging code.
The tool did not contribute to the research design, model development, or interpretation of the results, and did not influence the key arguments or conclusions.
All AI-assisted text and code were reviewed, verified, and edited by the authors, who take full responsibility for the content of this publication.

\section{Results and Discussion}
\label{sec_rad}

We benchmark EVE against the Vignes equation (Eq.~\eqref{eq_Vig}).
We also evaluated the Darken model~\cite{Darken1948}, the generalized Darken model~\cite{Krishna2005}, and the model of Bosse and Bart~\cite{Bosse2006} against the same test data.
The Darken and Bosse–Bart models are less accurate than the Vignes equation, and the generalized Darken model, while essentially as accurate, requires self-diffusion coefficients for the pure components as additional experimental input.
The Vignes equation is therefore the appropriate benchmark: among the reference models that require no input beyond the infinite-dilution diffusion coefficients, which EVE also uses, it is the most accurate.
The results for the other models are given in Figures S.2 and S.3 in the Supporting Information.

The comparison is carried out for Maxwell–Stefan diffusion coefficients.
As set out above, the values $\DMS^\mathrm{exp}_{ij}$ serving as references were derived from measured Fickian coefficients using thermodynamic factors from HANNA, which is also used within EVE.
This does not bias the comparison: the same $\Gamma_{ij}$ enters the conversion of the experimental data and the conversion of the predictions back to Fickian coefficients, so that the relative deviations reported below are identical for $\DMS^\mathrm{exp}_{ij}$ and $D^\mathrm{exp}_{ij}$, and hence hold equally for the directly measured quantity.
HANNA does, however, supply the activity-coefficient information used by the DSM, so that the results are not decoupled from HANNA. 

\subsection{Evaluation of EVE}

\begin{figure}
\centering
    \includegraphics[width=9cm]{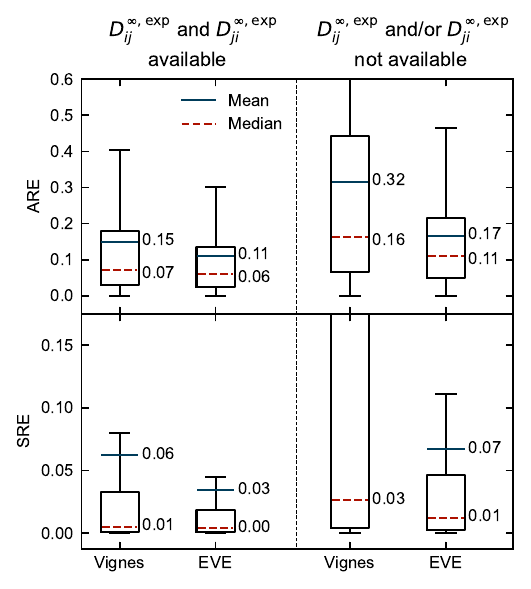}
    \caption{Boxplots of the absolute relative error~(ARE) and squared relative error~(SRE) of the predicted  Maxwell--Stefan diffusion coefficients~$\DMS_{ij}$ from the Vignes model (Eq.~\eqref{eq_Vig}) and EVE. The two columns show results for different sets of mixtures. Left: results for $N = 608$ data points of mixtures with both $D^{\infty, \mathrm{exp}}_{ij}$ and $D^{\infty, \mathrm{exp}}_{ji}$ available. Right: results for $N = 1498$ data points of mixtures for which at least one of them is unavailable, and the corresponding HADES prediction is used instead. Whisker length is 1.5 times the interquartile range; outliers are omitted for clarity. In the lower-right panel, the Vignes model's mean SRE (0.67) lies outside the displayed range.}
    \label{fig_BBV}
\end{figure}

Figure~\ref{fig_BBV} compares the Maxwell--Stefan diffusion coefficients, $\DMS_{ij}$, predicted by the Vignes model (Eq.~\eqref{eq_Vig}) and EVE with experimental data, in terms of ARE (top) and SRE (bottom).
The results are shown separately for two groups of mixtures.
For the left panels, both infinite-dilution diffusion coefficients are available from experiments; for the right panels, at least one had to be predicted with HADES.
The two groups thus correspond to the two situations in which EVE is applied in practice, and they comprise different mixtures, so that the panels are not directly comparable.
EVE improves on the Vignes model for both metrics in both cases.

For the first group (both infinite-dilution diffusion coefficients available, left), the mean ARE falls from 0.15 to 0.11 and the mean SRE from 0.06 to 0.03 when using EVE.
The reduction is larger for the mean than for the median, which decreases only from 0.07 to 0.06 in ARE, indicating that EVE improves on the Vignes equation mainly by reducing the number of severely mispredicted points rather than by a uniform shift.

For the second group (at least one infinite-dilution diffusion coefficients missing, right), the difference between the two models is considerably larger.
With EVE, the mean ARE falls from 0.32 to 0.17 and the median from 0.16 to 0.11; the mean SRE decreases from 0.67 to 0.07 and the median from 0.03 to 0.01.
The error distributions obtained with EVE are also markedly tighter, again reflecting the smaller number of poorly predicted points.

\FloatBarrier

\begin{figure}
\centering
    \includegraphics[width=15cm]{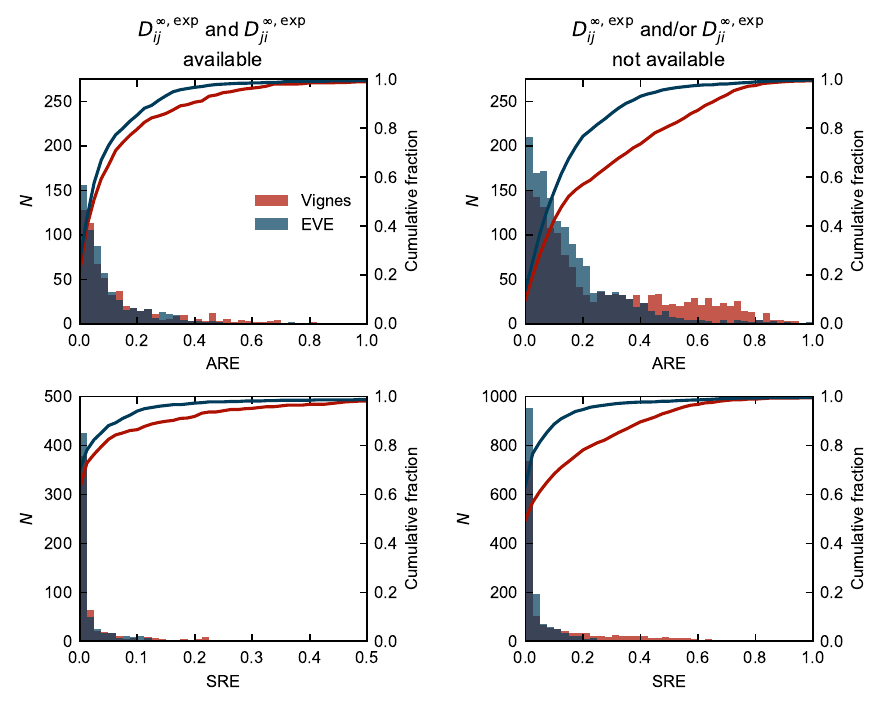}
    \caption{Histograms (bars) and cumulative fractions (lines) showing the number of $\DMS_{ij}$ predicted with a certain ARE (top row) and SRE (bottom row) with the Vignes model (Eq.~\eqref{eq_Vig}) and EVE. Left column: results for $N = 608$ data points of mixtures with both $D^{\infty, \mathrm{exp}}_{ij}$ and $D^{\infty, \mathrm{exp}}_{ji}$ available. Right column: results for $N = 1498$ data points of mixtures for which at least one of $D^{\infty, \mathrm{exp}}_{ij}$ and $D^{\infty, \mathrm{exp}}_{ji}$ is unavailable, and the corresponding HADES prediction is used instead.}
    \label{fig_Cum}
\end{figure}

Figure~\ref{fig_Cum} shows the same results in histograms, together with the cumulative fractions of test data points predicted with an error below a given value.
The two models differ not only in accuracy but in the shape of their error distributions.
For the first group (left panel), both models place a similar number of predictions in the lowest ARE and SRE bins ($\leq 0.05$), showing that a subset of these mixtures follows Vignes behavior closely.
Above these bins, however, the cumulative fractions of EVE exceed those of the Vignes model, so that EVE predicts fewer points with large errors.
For the second group (right panel), the Vignes model errors shift to distinctly larger values, and its cumulative fractions rise only gradually over a broad error range. In contrast, EVE outperforms it across the entire ARE and SRE range.

\FloatBarrier

\begin{figure}
\centering
    \includegraphics[width=15cm]{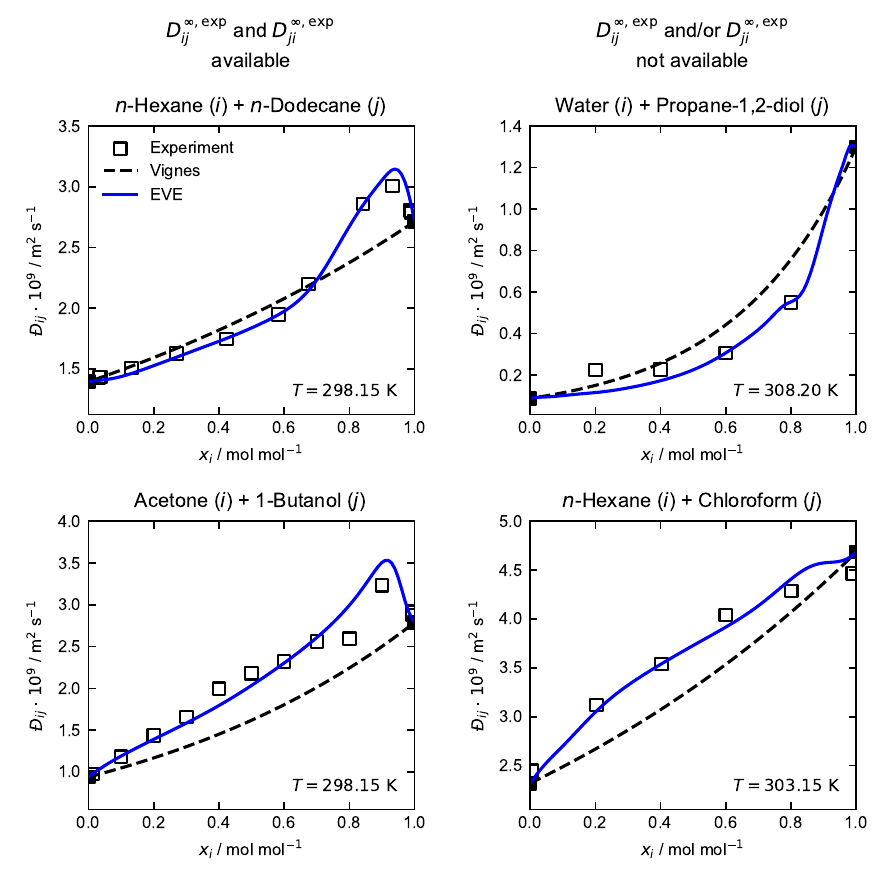}
    \caption{Maxwell--Stefan diffusion coefficients~$\DMS_{ij}$ as a function of the mole fraction~$x_i$ for two binary mixtures with both $D_{ij}^{\infty,\mathrm{exp}}$ and $D_{ji}^{\infty,\mathrm{exp}}$ available (left column) and two binary mixtures for which at least one of $D_{ij}^{\infty,\mathrm{exp}}$ and $D_{ji}^{\infty,\mathrm{exp}}$ is unavailable, and the corresponding HADES predictions are used instead (right column). Unfilled symbols represent experimental test data, while lines denote predictions from the Vignes model (Eq.~\eqref{eq_Vig}) and EVE. Filled symbols represent the infinite-dilution diffusion coefficients used as model inputs.}
    \label{fig_Exp}
\end{figure}

Figure~\ref{fig_Exp} shows $\DMS_{ij}$ as a function of composition for four exemplary mixtures, two from each group.
For (\textit{n}-hexane + \textit{n}-dodecane), from the first group, both models agree with the test data up to an \textit{n}-hexane mole fraction of about 0.7.
At higher \textit{n}-hexane mole fractions, the data depart from Vignes behavior, so that the Vignes model deviates increasingly, whereas EVE predicts the composition dependence correctly.
For (acetone + 1-butanol), also from the first group, the Vignes model misses the composition dependence over nearly the entire range, whereas EVE predicts it accurately.
The same pattern is found for (water + propane-1,2-diol) and (\textit{n}-hexane + chloroform), both from the second group: EVE is substantially more accurate than the Vignes model over most of the composition range, with larger deviations only at the lowest water and the highest \textit{n}-hexane mole fractions, where both models behave similarly.
These deviations likely stem in part from errors in the predicted infinite-dilution coefficients of water in propane-1,2-diol and chloroform in \textit{n}-hexane, which both models use as input.

\FloatBarrier

The behavior near the pure components deserves comment.
By construction, Eq.~\eqref{eq_EVE} suppresses the augmentation as either mole fraction approaches unity, so that EVE returns exactly the infinite-dilution coefficient there.
With $\alpha = 12$, this suppression sets in at roughly 0.1~mol~mol$^{-1}$ from the pure component.
Within this range, the damping function, rather than the augmentation, shapes the predicted curve: it interpolates between the infinite-dilution coefficient at the pure component and the augmented value at the edge of the damped region.
This is visible in Figure~\ref{fig_Exp} for $x_i$ above about 0.9~mol~mol$^{-1}$ for both (\textit{n}-hexane + \textit{n}-dodecane) and (acetone + 1-butanol), and to a lesser degree for (\textit{n}-hexane + chloroform).
This effect follows from enforcing the infinite-dilution boundary condition in an otherwise flexible model; by construction, it is confined to a narrow composition range and does not affect predictions elsewhere.
Figure~\ref{fig_Exp} also shows that the EVE predictions, while generally smooth, can exhibit slight undulations that the data do not resolve, as for (water + propane-1,2-diol) near $x_i = 0.8$~mol~mol$^{-1}$.
Such features reflect the flexibility of the augmentation in composition ranges where measurements are sparse.
They are small compared with the Vignes model deviations and do not appreciably affect prediction quality.

The practical value of EVE lies not only in its lower average error but also in its error distribution.
As Figure~\ref{fig_Cum} shows, the Vignes model performs excellently for some mixtures and poorly for others, and the applicable case cannot be determined without the very data the prediction is meant to replace.
EVE removes this uncertainty: it retains the Vignes model's accuracy where it works and avoids its large errors where it does not.
For the intended application, a mixture for which no experimental diffusion data exist, this is decisive, since the alternative is a prediction whose reliability is unknown.

\subsection{Fully Predictive Application of the Framework}
\label{sec_fle}

\begin{figure}
\centering
    \includegraphics[width=16.5cm]{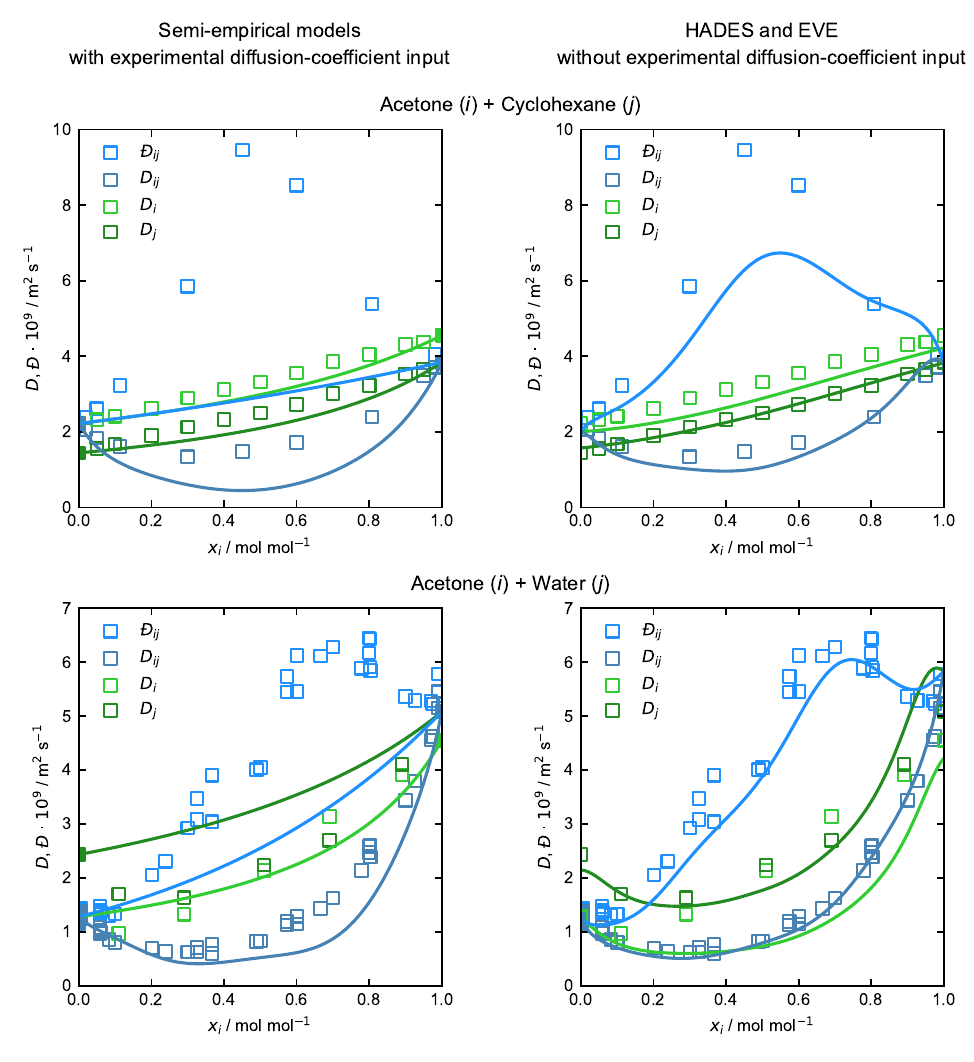}
    \caption{Maxwell--Stefan and Fickian mutual diffusion coefficients, $\DMS_{ij}$ and $D_{ij}$, and self-diffusion coefficients, $D_i$ and $D_j$, as functions of the mole fraction~$x_i$ for two binary mixtures (rows) at $T=298.15$~K. Unfilled symbols represent experimental test data, while lines denote predictions from the semi-empirical Vignes and Liu~\cite{Liu2011} models (left column) and HADES and EVE (right column). In the left column, filled symbols indicate the diffusion coefficients of the pure mixture components used to apply the semi-empirical models. All predictions in the right column were obtained without using any data on diffusion coefficients for the studied mixtures or components.}
    \label{fig_Flx}
\end{figure}

Figure~\ref{fig_Flx} compares the complete set of diffusion coefficients of two binary mixtures --- the Maxwell--Stefan and Fickian mutual diffusion coefficients, $\DMS_{ij}$ and $D_{ij}$, and the self-diffusion coefficients of both components, $D_i$ and $D_j$ --- with predictions from semi-empirical models (left column) and from HADES and EVE (right column).
Neither HADES nor EVE was trained on any diffusion coefficient of these mixtures, at finite concentrations or at infinite dilution.

The two columns are not on an equal footing.
The semi-empirical models require experimental diffusion data of the mixture components as input: the Vignes model uses the experimental infinite-dilution coefficients $D^{\infty, \mathrm{exp}}_{ij}$ and $D^{\infty, \mathrm{exp}}_{ji}$, and the model of Liu et al.~\cite{Liu2011}, which is the most accurate semi-empirical model for self-diffusion coefficients in mixtures~\cite{GuevaraCarrion2016} and was also used for benchmarking HADES~\cite{Wagner2026c}, additionally requires the self-diffusion coefficients of the pure components, $D^\mathrm{pure,exp}_{i}$ and $D^\mathrm{pure,exp}_{j}$.
HADES and EVE use none of these; their only inputs are the SMILES strings of the two components and the pure-component viscosities.
\FloatBarrier

For (acetone + cyclohexane), the Liu model adequately describes the self-diffusion coefficients of both components. However, the Vignes model does not capture the composition dependence of $\DMS_{ij}$, and the resulting Fickian coefficients $D_{ij}$ deviate accordingly.
HADES and EVE predict all four coefficients considerably better, although $\DMS_{ij}$, and with it $D_{ij}$, is underestimated in the middle of the composition range.

For (acetone + water), the picture is similar.
The semi-empirical models reproduce only the self-diffusion coefficient of acetone adequately and do not capture the strong composition dependence of the remaining coefficients.
HADES and EVE describe the mutual diffusion coefficients and the self-diffusion coefficient of water much better, but underestimate the self-diffusion coefficient of acetone in the mixture.

A further feature of Figure~\ref{fig_Flx} is that the curves for $\DMS_{ij}$, $D_{ij}$, and the self-diffusion coefficient of the infinitely diluted component predicted by HADES and EVE meet at the infinite-dilution limits.
This follows from the framework's construction: EVE's Vignes part uses the infinite-dilution coefficients predicted by HADES, and Eq.~\eqref{eq_EVE} reduces to the Vignes equation there.
These three diffusion coefficients, obtained from two different models, therefore converge to a single value at each limit, as they must.

As noted above, the experimental Maxwell--Stefan coefficients shown in Figure~\ref{fig_Flx} were derived from the measured Fickian coefficients using the thermodynamic factor from HANNA, and the predicted Fickian coefficients were obtained from the predicted Maxwell--Stefan coefficients using the same thermodynamic factor.
The basic experimental information used here is, therefore, that for the experimental Fickian diffusion coefficient $D_{ij}$, and the comparison between predicted and experimental values for $\DMS_{ij}$ and for $D_{ij}$ rests on the same basis, and the relative deviations to the experimental data are identical for both diffusion coefficients.
This does not affect the self-diffusion coefficients, which are compared directly with measured values.

Overall, our framework predicts all four diffusion coefficients of both mixtures without any experimental diffusion data, and does so more accurately than the semi-empirical models that require such data.
Deviations remain in individual cases for both the mutual and the self-diffusion coefficients, and additional high-quality experimental data, particularly for chemically diverse mixtures with pronounced composition dependence, would be needed to improve on this.

\section{Data Availability and Reproducibility Statement}

The numerical data from Figures~\ref{fig_BBV} and \ref{fig_Cum} and Figures~S.1--S.3 are tabulated in the Supporting Information.
The numerical data from Figures~\ref{fig_Exp} and \ref{fig_Flx} are not tabulated, as they contain experimental data from the DDB~\cite{DDB2025}, which cannot be redistributed for licensing reasons.
For the same reason, the model predictions shown in these figures cannot be provided, as they use experimental data as input, from which these data could be reconstructed.
The components are listed in Table~S.1, so that the database can be compiled by anyone with DDB access.
The final EVE model is available on Zenodo (\href{https://doi.org/10.5281/zenodo.22848728}{\mbox{https://doi.org/10.5281/zenodo.22848728}}) and via MLPROP (\href{https://ml-prop.mv.rptu.de/}{\mbox{https://ml-prop.mv.rptu.de/}}), where HADES and HANNA are also available, so that the three models can be used together as in this work.
The full development, training, and evaluation procedure of EVE is given in Section~\nameref{sec_tae} and the Supporting Information; the reliability of the predictions was assessed by system-wise cross-validation, which reflects their predictive application to mixtures without experimental data.

\section{Conclusions}

In this work, we introduce EVE, a hybrid model for predicting mutual diffusion coefficients in binary liquid mixtures.
EVE augments the Vignes equation with a deep-set model that takes molecular descriptors and activity-coefficient information from HANNA and predicts Maxwell--Stefan diffusion coefficients, from which the Fickian coefficients follow through the thermodynamic factor.
Evaluated on test data for mixtures excluded from training, EVE is more accurate than the Vignes equation and other available semi-empirical models, particularly for mixtures with no experimental infinite-dilution coefficients.
Its main advantage is the near-absence of large errors, which the Vignes equation produces for some mixtures in ways that are difficult to anticipate.

Combined with HADES, EVE yields the complete set of diffusion coefficients of a binary mixture, i.e., self-, Maxwell--Stefan, and Fickian diffusion coefficients, over the entire composition range and at any temperature, from the SMILES strings of the components and their pure-component viscosities alone.
No experimental diffusion data are required.
The two models' predictions are consistent: they converge to a common value at each infinite-dilution limit, so the four types of diffusion coefficients form a coherent set rather than independent estimates.

The same hybrid principle now underlies the prediction of both self- and mutual diffusion: a physical model provides the structure and the limiting behavior, while a machine-learning model supplies the flexibility that the physical model lacks.
That this transfers from the Stokes--Einstein equation (in HADES) to the Vignes equation (in EVE), and from self- to mutual diffusion, suggests it is not specific to either problem.
Furthermore, the deep-set architecture is not restricted to two components, and extending the approach to multicomponent mixtures is an ongoing area of work.

The augmentation in EVE is empirical.
It is constrained where physics constrains it, namely at the infinite-dilution limits and with respect to permutation of the components, but it is otherwise highly flexible. The framework was developed for mixtures of non-ionic organic components and water with molar masses below 1000~g~mol$^{-1}$ and no atoms heavier than chlorine.
Predictions outside this domain are possible but are not recommended.
These limits reflect the available experimental data rather than the modeling approach; extending them would require substantially larger, more diverse, high-quality diffusion datasets.
Ionic species are not covered and would require reconsidering the model's physical basis.

Beyond the direct prediction of diffusion coefficients, the framework could be applied to inverse problems, in which properties of unknown components, their molar mass, for instance, are inferred from measured diffusion coefficients.
This would be of particular interest for poorly specified mixtures, for example in combination with NMR fingerprinting techniques~\cite{Jirasek2019, Specht2021, Specht2023a, Specht2023b, Specht2023c, Wagner2025, Wagner2026}.

The framework offers a route to the diffusion coefficients of a binary liquid mixture that requires nothing more than the structures of its components and their pure-component viscosities, which are known or readily estimated for essentially any mixture of practical interest.

We fully disclose the trained model, its code, and a directly accessible variant via the interactive web interface MLPROP~\cite{Hoffmann2025} (\href{https://ml-prop.mv.rptu.de/}{\mbox{https://ml-prop.mv.rptu.de/}}), and we welcome improvements.

\newpage

\begin{acknowledgement}

We gratefully acknowledge financial support by the Carl Zeiss Foundation in the project `Halocycles', by DFG in the frame of the Research Training Group `WERA' (project number 503479768), the Core Facility `LASE-MR' (project number 537627671) and the Emmy Noether Group of FJ (project number 528649696), and by the German Federal Ministry of Research, Technology, and Space in the frame of the Project `IReKon'. Model training was carried out on the high-performance computer Elwetrisch at RPTU under the grant RPTU-MLVT.

\end{acknowledgement}

\section*{Conflict of Interest Statement}

The authors declare no conflict of interest.

\begin{suppinfo}

Computation of thermodynamic factors and Maxwell--Stefan diffusion coefficients; list of mixture components included in the database; hyperparameter study; comparison with reference models.

\end{suppinfo}

\bibliography{achemso-demo}

\end{document}

\section{Computation of Thermodynamic Factors and Maxwell--Stefan Diffusion Coefficients}

The HANNA ensemble model used in this work was obtained by training ten individual models with different random initializations following the same training procedure, hyperparameter settings, and data split as in the original work~\cite{Specht2024}, and is provided together with the code for the EVE model at Zenodo (\href{https://doi.org/10.5281/zenodo.22848728}{\mbox{https://doi.org/10.5281/zenodo.22848728}}).

The composition derivatives of the activity coefficients predicted by HANNA, used for the calculation of the thermodynamic factors (cf. Eq.~(2)) and as input to EVE (cf. Figure~1), were computed using the auto-differentiation function `autograd' from PyTorch~\cite{Paszke2019}.

Maxwell--Stefan diffusion coefficients were calculated using Fickian diffusion coefficients from the Dortmund Data Bank (DDB)~\cite{DDB2025} and thermodynamic factors predicted by HANNA (cf. Eq.~(1)). All data points for which HANNA yielded negative thermodynamic factors were excluded. In addition, all data points with $\DMS_{ij} > 10^{-8}$~m$^{2}$~s$^{-1}$ were removed to limit the influence of outliers and erroneous data, as such values fall outside the typical range for liquid-phase diffusion~\cite{Taylor1993}.
In total, these criteria excluded 536 data points, yielding a final dataset of $N = 2106$ data points used to train and evaluate EVE.

\begin{figure}
\centering
    \includegraphics[width=17cm]{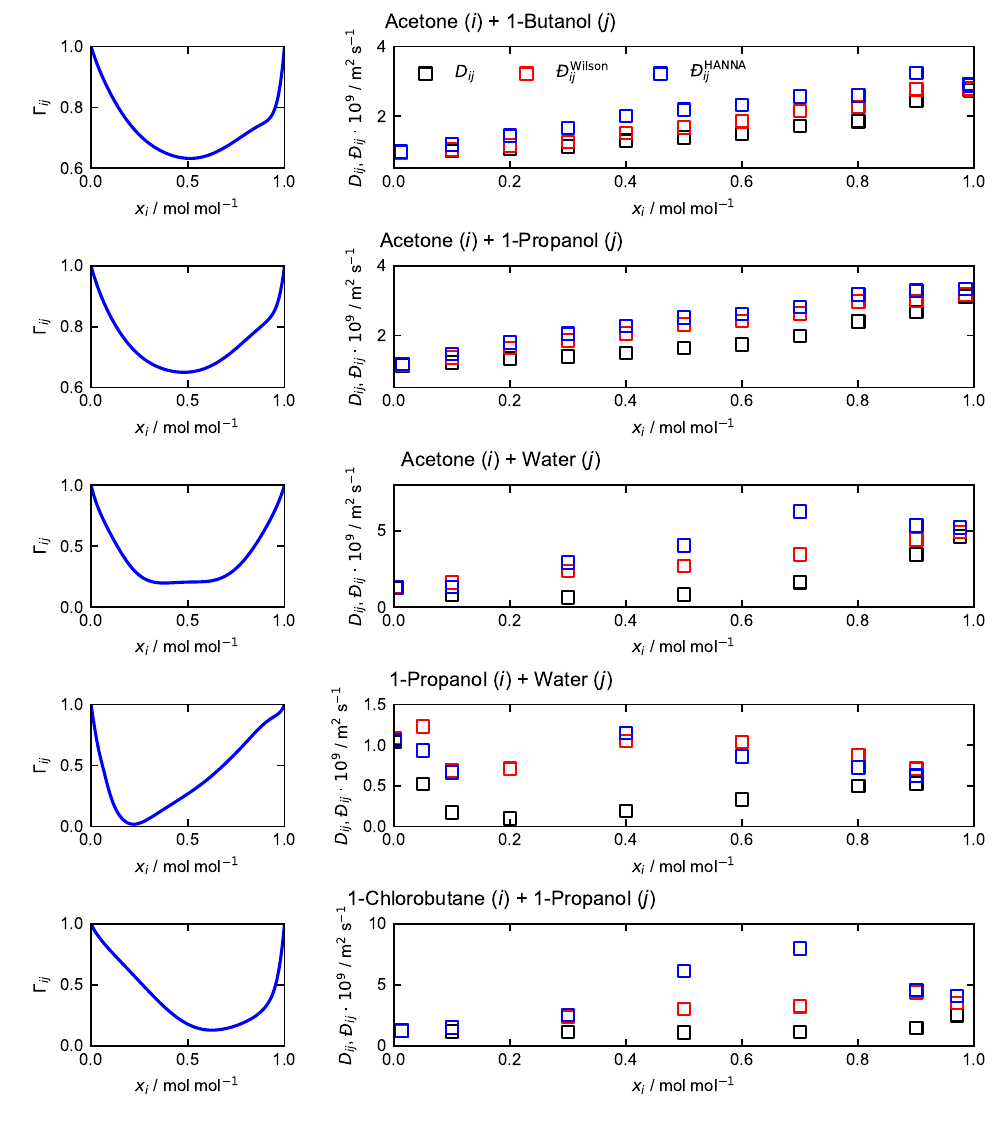}
    \caption{Thermodynamic factors~$\Gamma_{ij}$ from HANNA and resulting Maxwell--Stefan diffusion coefficients~$\DMS_{ij}$ as a function of the mole fraction~$x_i$ for five binary mixtures. Left: Thermodynamic factors from HANNA. Right: Comparison of the resulting calculated Maxwell--Stefan diffusion coefficients~$\DMS_{ij}^\mathrm{HANNA}$ with the underlying Fickian diffusion coefficients~$D_{ij}$ measured by Rehfeldt and Stichlmair~\cite{Rehfeldt2010} and the corresponding Maxwell--Stefan diffusion coefficients they calculated using the Wilson model~\cite{Taylor1991}~$\DMS_{ij}^\mathrm{Wilson}$.}
    \label{fig_HAN}
\end{figure}

Figure~\ref{fig_HAN} shows the thermodynamic factors~$\Gamma_{ij}$ predicted by HANNA and the resulting calculated Maxwell--Stefan diffusion coefficients~$\DMS_{ij}$ from this work as functions of the mole fractions $x_i$ for five binary mixtures, and compares them with the underlying Fickian diffusion coefficients $D_{ij}$ measured by Rehfeldt and Stichlmair~\cite{Rehfeldt2010} and the corresponding Maxwell--Stefan diffusion coefficients they calculated using the Wilson model~\cite{Taylor1991}. The results demonstrate that HANNA provides smooth and consistent results for $\Gamma_{ij}$ and the resulting $\DMS_{ij}$, with the deviation in the predicted diffusion coefficients from the reference values well within the variability observed when different activity-coefficient models are used~\cite{Taylor1991, Taylor1993}.

\FloatBarrier

\section{List of Mixture Components Included in the Database}

Table~\ref{tab_cmp} lists the names and SMILES strings of the mixture components included in the database of mutual diffusion coefficients used in the development of the EVE model.

\begin{longtable}{
    >{\raggedright\arraybackslash}p{0.52\textwidth}
    >{\raggedright\arraybackslash}p{0.42\textwidth}
}
\caption{Names and SMILES strings of the mixture components included in the database of mutual diffusion coefficients used in this work.}
\label{tab_cmp}\\

\toprule
Name & SMILES \\
\midrule
\endfirsthead

\multicolumn{2}{l}{Table~\thetable{} continued.}\\
\toprule
Name & SMILES \\
\midrule
\endhead

\bottomrule
\endlastfoot

(\textit{R})-2,3-Dimethylpentane & CC[C@@H](C)C(C)C \\
1,2-Ethanediol & OCCO \\
1,3,5-Triisopropylbenzene & CC(C)c1cc(C(C)C)cc(C(C)C)c1 \\
1,4-Dioxane & C1COCCO1 \\
12-Crown-4 & C1COCCOCCOCCO1 \\
15-Crown-5 & C1COCCOCCOCCOCCO1 \\
1-Butanol & CCCCO \\
1-Chlorobutane & CCCCCl \\
1-Methylnaphthalene & Cc1cccc2ccccc12 \\
1-Propanol & CCCO \\
2,2,4-Trimethylpentane & CC(C)CC(C)(C)C \\
2,3-Dimethylbutane & CC(C)C(C)C \\
(2\textit{R})-2-[(2\textit{R})-2-Hydroxypropoxy]propan-1-ol & C[C@H](CO)OC[C@@H](C)O \\
2-Butanone & CCC(C)=O \\
2-Methyl-1-propanol & CC(C)CO \\
2-Propanol & CC(C)O \\
3-Methylpentane & CCC(C)CC \\
Acetic acid & CC(=O)O \\
Acetone & CC(C)=O \\
Acetonitrile & CC\#N \\
Aniline & Nc1ccccc1 \\
Benzene & c1ccccc1 \\
Carbon disulfide & S=C=S \\
Carbonic acid dimethyl ester & COC(=O)OC \\
Chloroform & ClC(Cl)Cl \\
Cyclohexane & C1CCCCC1 \\
Cyclohexanone & O=C1CCCCC1 \\
Diethanolamine & OCCNCCO \\
Diethyl carbonate & CCOC(=O)OCC \\
Diethyl ether & CCOCC \\
Diethyl oxalate & CCOC(=O)C(=O)OCC \\
Diethylene glycol & OCCOCCO \\
Diisopropyl ether & CC(C)OC(C)C \\
Ethanol & CCO \\
Ethyl acetate & CCOC(C)=O \\
Ethylbenzene & CCc1ccccc1 \\
Geranial & CC(C)=CCC/C(C)=C/C=O \\
Hexafluorobenzene & Fc1c(F)c(F)c(F)c(F)c1F \\
Mesitylene & Cc1cc(C)cc(C)c1 \\
Methanol & CO \\
Methyl isobutyl ketone & CC(=O)CC(C)C \\
Methyl isopropyl ketone & CC(=O)C(C)C \\
Methylcyclohexane & CC1CCCCC1 \\
\textit{n}-Butyl acetate & CCCCOC(C)=O \\
\textit{n}-Decane & CCCCCCCCCC \\
\textit{n}-Dodecane & CCCCCCCCCCCC \\
\textit{N}-Formylmorpholine & O=CN1CCOCC1 \\
\textit{n}-Heptane & CCCCCCC \\
\textit{n}-Hexadecane & CCCCCCCCCCCCCCCC \\
\textit{n}-Hexane & CCCCCC \\
\textit{N}-Methyl-2-pyrrolidone & CN1CCCC1=O \\
\textit{N}-Methyldiethanolamine & CN(CCO)CCO \\
\textit{N},\textit{N}-Dimethylformamide & CN(C)C=O \\
\textit{n}-Octane & CCCCCCCC \\
\textit{n}-Tetradecane & CCCCCCCCCCCCCC \\
\textit{o}-Xylene & Cc1ccccc1C \\
Phenol & Oc1ccccc1 \\
(\textit{R})-Propane-1,2-diol & C[C@@H](O)CO \\
Sulfolane & O=S1(=O)CCCC1 \\
\textit{tert}-Butanol & CC(C)(C)O \\
Tetrachloromethane & ClC(Cl)(Cl)Cl \\
Tetralin & c1ccc2c(c1)CCCC2 \\
Toluene & Cc1ccccc1 \\
\textit{trans}-Decalin & C1CC[C@H]2CCCC[C@@H]2C1 \\
Trifluoroacetic acid & O=C(O)C(F)(F)F \\
Water & O \\

\end{longtable}
\FloatBarrier

\section{Hyperparameter Study}

Table~\ref{tab_hyper} summarizes the hyperparameters of EVE varied in this work, namely the AdamW optimizer weight decay $\lambda$, the initial learning rate, the batch size, the number of layers before and after the deep-set layer~\cite{Zaheer2017}, the number of nodes per DSM layer, and the EVE hyperparameter $\alpha$~(cf. Eq.~(4)), together with the corresponding validation loss (MSRE) obtained for each model configuration. We chose the optimal hyperparameter configuration based on the lowest validation loss and model complexity, and flexibility, which increases with $\alpha$. We therefore selected Model~1 and used it throughout the manuscript.

\begin{table}
\centering
\caption{Tested EVE model variants with varied hyperparameters and the corresponding validation loss. $\lambda$ denotes the weight decay of the AdamW optimizer and $\alpha$ the EVE hyperparameter.}
\label{tab_hyper}
\begin{adjustbox}{width=\textwidth}
    \begin{tabular}{llllllll}
        \toprule
        Model No. & $\lambda$ & Initial learning rate & Batch size & Layers & Nodes & $\alpha$ & Validation loss\\
        \midrule
        1  & $5 \cdot 10^{-3}$ & 0.001 & 1  & 1 & 32 & 12 &  0.0332 \\
        2  & $1 \cdot 10^{-4}$ & 0.001 & 1  & 1 & 32 & 12 &  0.0382 \\
        3  & $1 \cdot 10^{-3}$ & 0.001 & 1  & 1 & 32 & 12 &  0.0372 \\
        4  & $5 \cdot 10^{-3}$ & 0.0001 & 1  & 1 & 32 & 12 &  0.0332 \\
        5  & $5 \cdot 10^{-3}$ & 0.01 & 1  & 1 & 32 & 12 &  0.0332 \\
        6  & $5 \cdot 10^{-3}$ & 0.001 & 2  & 1 & 32 & 12 &  0.0462 \\
        7  & $5 \cdot 10^{-3}$ & 0.001 & 4  & 1 & 32 & 12 &  0.0512 \\
        8  & $5 \cdot 10^{-3}$ & 0.001 & 1  & 2 & 32 & 12 &  0.0392 \\
        9  & $5 \cdot 10^{-3}$ & 0.001 & 1  & 3 & 32 & 12 &  0.0772 \\
        10  & $5 \cdot 10^{-3}$ & 0.001 & 1  & 1 & 16 & 12 &  0.0392 \\
        11  & $5 \cdot 10^{-3}$ & 0.001 & 1  & 1 & 64 & 12 &  0.0372 \\
        12  & $5 \cdot 10^{-3}$ & 0.001 & 1  & 1 & 32 & 5 &  0.0420 \\
        13  & $5 \cdot 10^{-3}$ & 0.001 & 1  & 1 & 32 & 8 &  0.0332 \\
        
        \bottomrule
    \end{tabular}
\end{adjustbox}
\end{table}
\FloatBarrier

\section{Comparison with Reference Models}

In the following, the performance of reference models~\cite{Darken1948, Krishna2005, Bosse2006} for predicting Maxwell--Stefan diffusion coefficients~$\DMS_{ij}$ in binary mixtures, which are listed in Table~\ref{tab_ref} together with their required inputs, is compared to that of the Vignes model~\cite{Vignes1966}, which serves as the primary reference in the manuscript, and of EVE using the test data from the present work.
\begin{sidewaystable} % rotates entire float incl. caption by 90°
\centering
\begin{threeparttable} % caption and notes take the tabular's width
\caption{Reference models for predicting Maxwell--Stefan diffusion coefficients~$\DMS_{ij}$ in binary mixtures and the inputs they require.}
\label{tab_ref}
    \begin{tabular}{lll}
        \toprule
        Model & & Required input\\
        \midrule
        Darken~\cite{Darken1948}  
        & $\DMS_{ij} = x_i D_j + x_j D_i$
        & $D_i$, $D_j$\tnote{a}\\
        Generalized Darken~\cite{Krishna2005}
        & $\DMS_{ij} = x_i \left( x_i^{(m)} D^\infty_{ji} +  x_j^{(m)} D^\mathrm{pure}_{j} \right) 
        + x_j \left( x_i^{(m)} D^\mathrm{pure}_{i}  +  x_j^{(m)} D^\infty_{ij}\right)$\tnote{b}
        & $D^\infty_{ij}$, $D^\infty_{ji}$, $D^\mathrm{pure}_{i}$, $D^\mathrm{pure}_{j}$\\
        Bosse and Bart (BB)~\cite{Bosse2006}  
        & $\DMS_{ij} = \left( D_{ij}^{\infty} \right)^{x_j} 
        \left( D_{ji}^{\infty} \right)^{x_i} 
        \exp \left( -\frac{g^\mathrm{E}}{RT} \right)$
        & $D^\infty_{ij}$, $D^\infty_{ji}$, $g^\mathrm{E}$\\
        \bottomrule
    \end{tabular}
    \begin{tablenotes}
        \footnotesize
        \item[a] $D_i$ and $D_j$ are the self-diffusion coefficients of components $i$ and $j$ in the mixture, in contrast to the pure-component values $D^\mathrm{pure}_{i}$ and $D^\mathrm{pure}_{j}$.
        \item[b] $x_i^{(m)}$ and $x_j^{(m)}$ are the mass fractions of components $i$ and $j$, respectively.
    \end{tablenotes}
\end{threeparttable}
\end{sidewaystable}
\FloatBarrier

\begin{figure}
\centering
    \includegraphics[width=12cm]{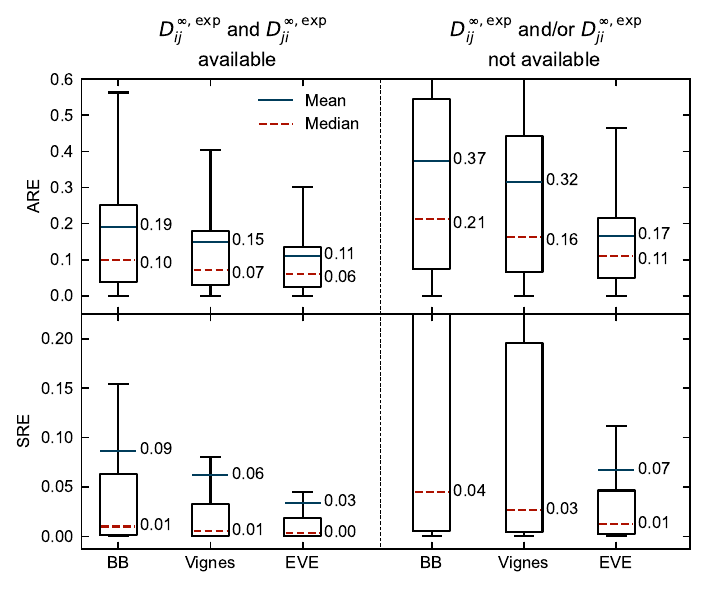}
    \caption{Boxplots of the absolute relative error~(ARE) and squared relative error~(SRE) of the predicted Maxwell--Stefan diffusion coefficients~$\DMS_{ij}$ from the Bosse and Bart (BB)~\cite{Bosse2006} model, the Vignes model and EVE. Left: results for $N = 608$ data points of mixtures with both $D^{\infty, \mathrm{exp}}_{ij}$ and $D^{\infty, \mathrm{exp}}_{ji}$ available. Right: results for $N = 1498$ data points of mixtures for which at least one of them is unavailable, and the corresponding HADES prediction is used instead. The whisker length is 1.5 times the interquartile range; outliers are omitted for visual clarity. In the lower right panel, the mean SRE of the BB model (0.86) and the Vignes model (0.67) lie outside the displayed range.}
    \label{fig_BBB}
\end{figure}

Figure~\ref{fig_BBB} shows the ARE (top) and SRE (bottom) calculated by comparing the predicted $\DMS_{ij}$ from the model of Bosse and Bart (BB)~\cite{Bosse2006}, using the excess Gibbs energy~$g^\mathrm{E}$ derived from HANNA here (cf. Table~\ref{tab_ref}), to the experimental test data in boxplots, and compares them with those of the Vignes model and EVE reported in the manuscript.
The left panel shows mixtures for which the experimental infinite-dilution data, $D_{ij}^{\infty,\mathrm{exp}}$ and $D_{ji}^{\infty,\mathrm{exp}}$, are available ($N = 608$); the right panel shows those for which at least one of them is missing and replaced by the corresponding HADES prediction ($N = 1498$).
The results show that the BB model performs clearly worse than the Vignes baseline in both cases.
\FloatBarrier

The Darken model~\cite{Darken1948} requires self-diffusion coefficients in the mixture, and the generalized Darken model~\cite{Krishna2005} requires those of the pure components; for both, only limited experimental data are available~\cite{DDB2025, Wagner2026c}.
To enable a consistent comparison with the Vignes model and EVE across the full database, all models are evaluated using HADES predictions exclusively as inputs, so that the distinction made in Figure~\ref{fig_BBB} does not apply here and all $N = 2106$ data points are shown together.

\begin{figure}
\centering
    \includegraphics[width=8.75cm]{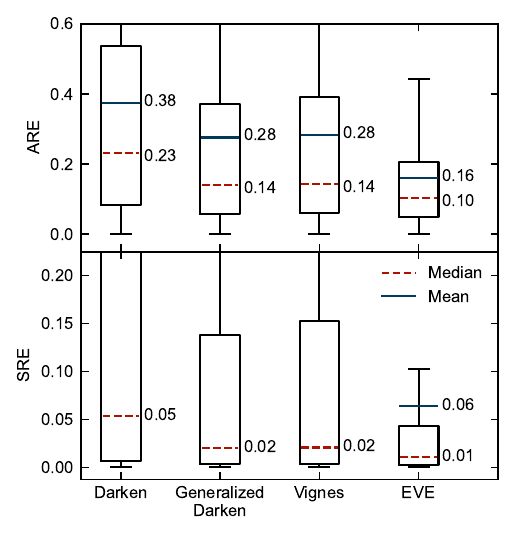}
    \caption{Boxplots of the absolute relative error~(ARE) and squared relative error~(SRE) of the predicted  Maxwell--Stefan diffusion coefficients~$\DMS_{ij}$ from the Darken model~\cite{Darken1948}, the generalized Darken model~\cite{Krishna2005}, the Vignes model, and EVE using HADES predictions as inputs. The whisker length is 1.5 times the interquartile range; outliers are omitted for visual clarity. In the lower panel, the mean SREs of the Darken (0.84), generalized Darken (0.59), and Vignes (0.53) models lie outside the displayed range.}
    \label{fig_BBR}
\end{figure}

Figure~\ref{fig_BBR} shows the ARE (top) and SRE (bottom) obtained by comparing the predicted $\DMS_{ij}$ to the experimental test data, presented as boxplots. The results show that the Darken model performs worse than the Vignes baseline, while the generalized Darken model shows comparable predictive accuracy. However, because the generalized Darken model requires self-diffusion coefficients for the pure mixture components, which prevents extensive assessment using experimental data~\cite{DDB2025, Wagner2026c}, we selected the Vignes model as the benchmark throughout the manuscript.
\FloatBarrier

\newpage

\bibliography{achemso-demo}